# Phase noise characterization of sub-hertz linewidth lasers via digital cross correlation

XIAOPENG XIE,[1] ROMAIN BOUCHAND,[1] DANIELE NICOLODI,[1,+] MICHEL LOURS,[1] CHRISTOPHE ALEXANDRE,[2] AND YANN LE COQ[1,*]

[1]LNE-SYRTE, Observatoire de Paris, PSL Research University, CNRS, Sorbonne Universités, UPMC Univ. Paris 06, 61 avenue de l'Observatoire, 75014 Paris, France
[2] CNAM, CEDRIC Laboratory, 292 rue Saint Martin, 75003 Paris, France
[+]Current address: National Institute of Standards and Technology (NIST), 325 Broadway, Boulder, Colorado 80305, USA
*Corresponding author: yann.lecoq@obspm.fr

**Phase noise or frequency noise is a key metrics to evaluate the short term stability of a laser. This property is of a great interest for the applications but delicate to characterize, especially for narrow line-width lasers. In this letter, we demonstrate a digital cross correlation scheme to characterize the absolute phase noise of sub-hertz line-width lasers. Three 1,542 nm ultra-stable lasers are used in this approach. For each measurement two lasers act as references to characterize a third one. Phase noise power spectral density from 0.5 Hz to 0.8 MHz Fourier frequencies can be derived for each laser by a mere change in the configuration of the lasers. This is the first time showing the phase noise of sub-hertz line-width lasers with no reference limitation. We also present an analysis of the laser phase noise performance.**

Laser phase noise describes how the phase of a laser output electrical field deviates from an ideal sinusoidal wave. This quantity which is defined to evaluate the short term stability of a laser can also be used to estimate the line-width or coherent length of a laser. In many applications, such as coherent optical communication [1], LIDAR [2], optical fiber-based interferometeric sensors [3], high resolution spectroscopy [4], ultra-low phase noise photonic microwave generation [5], and optical atomic clock [6], the laser phase noise can profoundly impacts the limitation of a system. Thus, lasers with ultra-low phase noise are actively studied [7-10]; while the precise characterization of such ultra-stable laser is becoming more important.

Phase noise characterization is a comparison process. Generally, laser phase noise measurement approaches can be divided into two categories according to the comparison method. The first one is comparing the laser under test with itself through the schemes of delayed self-homodyne, delayed self-heterodyne [11,12] or Michelson interferometer [13]. Several kilometers of fiber are usually used for these optical delay line methods in order to make the delay time longer than the laser coherent time. It is difficult to characterize the phase noise of lasers with hundreds-hertz line-width or narrower via these delay line technique as thousands of kilometers of fiber would be required, and the fiber noise itself can also become a limitation. The second category, which is called the beat-note method [14-17], implies comparing the laser under test with a reference laser whose phase noise is much lower than that of the one under test. When the phase noise of the laser under test is lower than that of any available reference, two similar lasers must be built and compared. Assuming statistical independence and equal contribution of both lasers, the phase noise is revealed after division of the beat-note phase noise by $\sqrt{2}$. However, to realize two equally good lasers is not straightforward.

Cross correlation is a well know approach to characterize the phase noise of RF and microwave oscillators with ultra-low level [5, 18-22]. Here, we extend this approach to precisely characterize the phase noise of sub-hertz line-width lasers. Three ultra-stable 1,542 nm lasers are introduced in our system. Two of them act as references and beat with the one under test to generate two radio frequency electrical signals after photo-detection. These two beat notes are analyzed by a home-designed digital heterodyne cross correlator [5, 21]. Both of the electrical beat-note signals carry the phase information of the laser under test. As noise contributions from the two other optical sources are statistically independent, the phase noise power spectral density (PSD) of the laser under test is revealed by averaging the statistical estimator of the cross PSD of the phases of these two beat note signals. Unlike the traditional beat-note phase noise characterization, the reference lasers in this cross correlation system do not need to possess better phase noise level than that of the laser under test. Note that this is, in a sense, similar with the so called three cornered hat method [23]. However our system provides the full phase noise PSD of the laser under test. Furthermore in this work, we switch the role of the three lasers so as to characterize the absolute phase noise of each of them. Phase noise PSDs from 0.5 Hz to 0.8 MHz

Fourier frequencies are shown for these three lasers. Besides, the phase noise performance is analyzed.

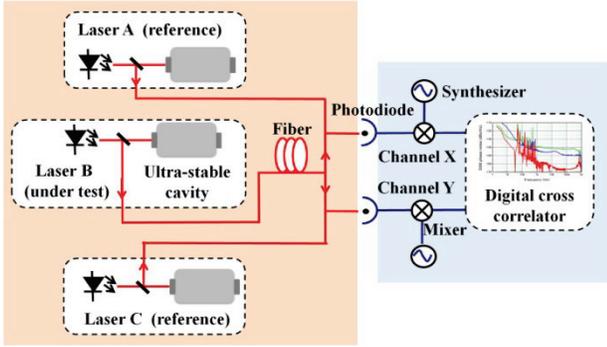

**Fig. 1.** Experiment set-up of the laser phase noise characterization. Three separate ultra-stable 1,542 nm lasers are used in this scheme. The laser B is beat with the reference laser A and C to get two electrical signals. These two electrical signals are first mixed-down to a lower frequency and then analyzed by a home-designed digital cross-correlator to reveal the phase noise PSD of the laser B.

Fig. 1 shows our cross-correlation scheme to characterize the phase noise of narrow line-width lasers. These three lasers achieve narrow line-width regime through active stabilization of semiconductor laser diode to ultra-stable cavity by the Pound-Drever-Hall (PDH) method [24]. As these three lasers are located in two different rooms, their output lights are transmitted through tens of meters of fiber to the optical table where the photodiode is settled. The fiber links are noise-cancelled by the acousto-optic modulator based feedback [25]. Fiber polarization controllers are used on each reference laser branch to reach sufficient beat note signal power. The frequency differences between each pair of lasers are within 600 MHz.

To precisely characterize the phase noise of the laser B (under test), it is beat with two distinct reference lasers (A and C), as is presented in Fig.1. These beat notes are separately detected by fast InGaAs photodiodes in channel X and Y. The output power of the photodiode is below -40 dBm. RF amplifier with 20 dB gain is used after the photodiode. The first stage of our home-designed digital cross correlator is analog to digital conversion (ADC) at 250 mega sample per second (MSPS), which requires input frequencies below 125 MHz to obey Nyquist criterion. A frequency synthesizer therefore acts as the local oscillator to down-convert the carrier frequency of the beat note signal to approximately 10 MHz (the two channels X and Y do not need to have exactly the same frequency). The down-converted signals are power-amplified, low-pass filtered, and sent into two different ADCs and fed to a field-programmable-gate-array (FPGA) based digital cross correlator. The internal structure of this digital correlator is similar with we reported in [5, 21]. The digitized samples are demodulated in digital down conversion logic units, providing two phase data streams at 2 MSPS, which are then transferred by gigabit Ethernet link to a computer for easy data analysis. Data streams are frequency de-drifted before they are analyzed, which is to overcome the frequency drift of laser cavity during the measurement. As both of these phase data sets from channel X and channel Y carry the phase information of the laser under test, averaging the cross phase noise PSD $S_{YX}$ of these two data sets converges towards the phase noise PSD $S_{BB}$ of the laser under test [18]:

$$\begin{aligned}
\langle S_{YX} \rangle_m &= \frac{2}{T} \langle YX^* \rangle_m \\
&= \frac{2}{T} \langle [B+C] \times [B+A]^* \rangle_m \\
&= \frac{2}{T} [\langle BB^* \rangle_m + \langle BA^* \rangle_m + \langle CB^* \rangle_m + \langle CA^* \rangle_m] \\
&= S_{BB} + O(\sqrt{1/m}).
\end{aligned} \qquad (1)$$

where m is the average number, T is the measurement time. As three lasers are statistically independence, the cross terms decrease with a speed of 5log[m] dB during the averaging process until it reaches the final phase noise PSD of the laser under test. Note that reference lasers with levels of phase noise higher than that of the laser under test are possible but require a longer measurement time. For instance, if the reference lasers exhibit phase noise levels one order of magnitude higher than that of the laser under test, our system requires a longer time to converge to the final phase noise PSD of the laser under test (For example, it can take up to one hour for the first decade). In fact, these two time-dependent data sets from channel X and Y do not only take the noise from the reference laser, but also carry the noise of photodiode, local oscillator synthesizers, and the digital cross correlator. The noise added by the photodiode, the local oscillator synthesizers and the digital cross correlator are far below the phase noise of the lasers we characterize. Furthermore, and more importantly, they are also for a large part statistically independence, and thus are averaged out during the laser phase noise characterization.

Fig.2 presents the electrical spectra of the beat note signals that measured before the mixer. The carrier frequencies of these two beat notes after the photodiode are 555 MHz and 303 MHz respectively. From the electrical spectra of the beat note, we can infer that the line-widths (full width at half-maximum, FWHM) of all these lasers are narrower than 1 Hz, but this estimation is limited by the resolution bandwidth of the electrical spectrum analyzer. In Fig.2 (a), the symmetrical bump around 250 kHz offset frequency is due to one of the PDH lock (from laser under test or reference laser A). The same situation is observed in Fig.2 (b), but the locking bump is around 20 kHz. Combining the information from Fig.2 (a) and (b), the locking bump can be deduced to originate from the reference lasers but not the laser under test as the locking bump offset frequency are different in these two figures. Averaging the cross phase noise PSD of these beat-notes signal can give the individual laser under test information.

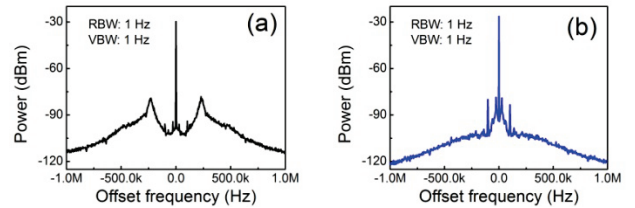

**Fig. 2.** Electrical spectra of the laser beat note signals. RBW: resolution band-width, VBW: video band-width. (a) Beat note between the laser B (under test) and reference laser A. (b) Beat note between laser B and reference laser C.

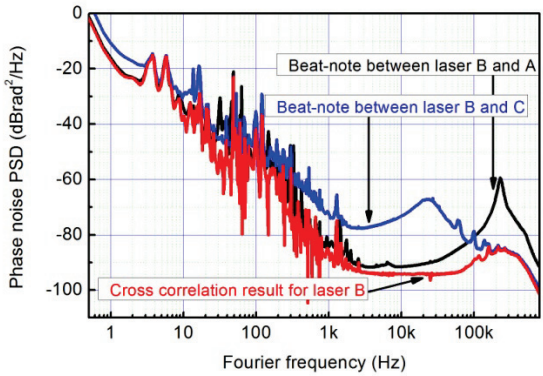

**Fig. 3.** Blue and black: phase noise PSDs for the beat notes of laser B (under test) and the two references lasers (A and C). Phase noise PSD of the beat note includes the contribution from both the laser under test and corresponding reference laser. Red: cross phase PSD result. This represents the absolute phase noise PSD of the laser under test, which is obtained by the heterodyne cross-correlator.

Fig.3 displays the typical result obtained for characterizing laser B by the cross correlator device. All phase noise PSD plots span to 0.8 MHz as the effect of low-pass filter. Black and blue curves corresponds to direct PSD of phase comparison between laser B and laser A (black) and laser C (blue). These curves are in very good agreement with the corresponding electrical spectra shown in Fig.2. In particular the PDH phase locking bumps at 250 kHz (black) and 20 kHz (blue) clearly correspond to the similar features seen on the spectra displayed on fig 2. The red curve corresponds to the cross phase PSD result, and therefore to the absolute phase noise PSD of laser B, assuming statistical independence between the phase fluctuations of A, B and C. In this curve it appears obvious that the previously observed servo bumps are coming from laser A and C, while the servo bump of laser B is much lower, and dwarfed in the A-B beat-note comparison by that of laser A. In the blue curve, on the contrary, the minor feature near 250 kHz is indeed coming from the servo bump of laser A, whereas the largest bump near 20 kHz is a feature of laser C alone

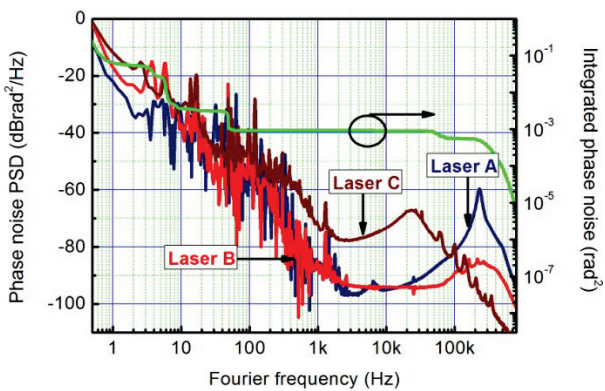

**Fig. 4.** Individual absolute phase noise PSDs of all three lasers used in the cross correlation scheme. The phase noise PSDs of the two reference lasers A and C are characterized just by switching their roles in the cross correlation scheme. Green curve is the integrated phase noise of laser B.

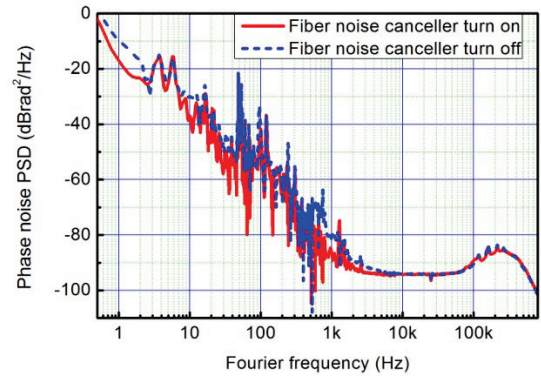

**Fig. 5**. Phase noise PSD of the narrow line-width laser is affected by the acoustic noise at low Fourier frequencies.

Furthermore, we can also observe that at low Fourier frequencies from 0.5 to 10 Hz, the phase noise of the laser B is higher than that of the reference laser A, thus limiting the phase noise PSD of the beat note between laser B and A. These slight differences between three lasers at different Fourier frequencies can be precisely characterized by the cross correlator system, which is one of the attractive features of this laser phase noise measurement method.

By switching the respective roles of the lasers in Fig.1, we can also iteratively characterize the phase noise of the other two lasers A and C. Fig.4 presents the final phase noise PSD results of all three lasers obtained this way. For all these lasers, the phase noise from 1 kHz to 1 MHz Fourier frequencies is limited by the PDH locking finite gain and bandwidth combined with the free running noise of the semiconductor lasers used in the servo loop. From looking at the locking bumps, it is obvious that the three lasers have different PDH locking bandwidths. This phase locking bandwidth could be improved by using lower noise lasers, and optimizing the loop filter circuit and feedback capability. Below 1 kHz Fourier frequencies, one observes some spurs that do not average down, even after several hours. They are partly caused by the 50 Hz and its harmonics from the various DC power supplies in the experiment, and partly by the acoustic noise of the fiber links that transmit the laser signals to the photodiodes. As shown in Fig.5, the fiber link noise can reject up to 10 dB of acoustically-induced phase noise for certain Fourier frequencies. The ultra-stable cavity systems are carefully enclosed in acoustically isolating boxes. However, even when the fibers connecting the various systems together are noise cancelled, there remains approximately 2 meters of fiber that could not be noise canceled right before the laser beat-note generation. We believe these un-noise compensated fibers to be the main remaining sources of acoustic noise in the system. Note that sometimes spurs on the two channels have negative correlation during the measurement, like the tiny spur at 25 kHz in Fig.4. The phase noise PSD at low Fourier frequencies is determined by the fractional frequency stability of the ultra-stable cavity to which the laser is locked. In our case, electrical noise contributions in the PDH locking make these phase noise curves depart from thermal noise limit of the ultra-stable cavity [26]. The phase noise PSD values of these three lasers at 1 Hz offset are comparable but slightly different. According to the relationship between the phase noise and frequency noise:

$$S_\nu(f) = f^2 S_\varphi(f) \qquad (2)$$

where $S_\varphi(f)$ is the phase noise PSD, $S_\nu(f)$ is the frequency noise PSD, $f$ is the Fourier frequency, we can convert the phase noise PSD in Fig.4 to frequency noise PSD very easily. Meanwhile, one possible the relation between FWHM line width and phase noise can be defined as [27]:

$$\int_{\frac{\Delta\nu}{2}}^{+\infty} S_\varphi(f) df = \frac{2}{\pi} \qquad (3)$$

where $\Delta\nu$ is the FWHM line width of the laser. From the phase noise measurement result, we can verify that, with this definition, all three lasers have line widths narrower than 1 Hz. This fact is exemplified in Fig. 4, where the integrated phase noise from high Fourier frequencies down to lower Fourier frequencies is plotted and is shown to remain substantially below $2/\pi$ until 0.5 Hz Fourier frequency and lower.

In summary, we have successfully characterized the phase noise of sub-hertz line-width lasers by using a FPGA-based heterodyne cross correlator. The noise floor of the home-made cross correlator is far below the phase noise of any reported ultra-narrow line-width laser [5]. In order to characterize the phase noise of an individual narrow line-width laser by cross-correlation, we have to use two additional reference lasers. The frequency difference between the laser under test and reference lasers needs here to remain within the bandwidth of the photodiode that convert the optical beat notes to electrical signals. Using an extra optical frequency comb could overcome this frequency bandwidth limitation. In a few words, provided the laser under test and the optical frequency comb provide a large enough optical beat signal to noise ratio, phase locking one of the optical frequency comb lines to the laser under test could transfer the spectral properties of the laser under test to each comb line, and thus offer a bandwidth equal to that of the optical frequency comb, i.e. hundreds of nanometers [28]. In this way, beating the reference lasers with the comb lines nearest to their carrier optical frequencies would produce electrical beat notes that carry information about the phase of the laser under test. Averaging the cross phase PSD of the two beat notes generated by different reference lasers, with an extra added mathematical step to normalize the various measured phases to the carrier frequency of the laser under test would produce the phase noise PSD of this laser. The demonstrated cross correlation method and its further developments will therefore be a useful tool to characterize the individual phase noise of extremely narrow line-width laser, regardless of their carrier optical frequencies.

**Funding.** Defense Advanced Research Projects Agency (DARPA) Program in Ultrafast Laser Science and Engineering (PµreComb project, under contract No. W31P4Q-14-C-0050); the Formation, Innovation, Recherche, Services et Transfert en Temps-Fréquence (FIRST-TF); Labex and the Eurostar Eureka program (Stable Microwave Generation and Dissemination over Optical Fiber project).

**Acknowledgement.** We thank José Pinto for help with the electronics and Rodolphe Le Targat for the laser distribution.